\documentclass[prl,twocolumn,showpacs,preprintnumbers,amsmath,amssymb]{revtex4}
\usepackage{graphicx}
\usepackage{dcolumn}
\usepackage{bm}
\usepackage{subfigure}

\begin{document}

\title{Convex Hull of $N$ Planar Brownian Motions: Exact Results and 
an Application to Ecology}
\author{Julien Randon-Furling$^1$}
\author{Satya N. Majumdar$^1$}
\author{Alain Comtet$^{1,2} $}
\affiliation{$^1$ Laboratoire de Physique Th\'{e}orique et Mod\`{e}les 
Statistiques (UMR CNRS 8626)\\ Universit\'{e} Paris-Sud, B\^{a}t. 100, 
91405 Orsay Cedex, France\\
$^2$ Institut Henri Poincar\'e, Universit\'e Pierre et Marie Curie-Paris 6, France.}

\date{\today}

\begin{abstract}
We compute exactly the mean perimeter and area 
of the convex hull of $N$ independent planar Brownian paths each
of duration $T$, both for open and closed paths. 
We show that the mean perimeter $\displaystyle \langle
L_N\rangle = \alpha_N\, \sqrt{T}$ and the mean area
$\displaystyle \langle A_N\rangle =
\beta_N\, T$ for all $T$. The prefactors $\alpha_N$ and $\beta_N$, computed exactly for all 
$N$,
increase very slowly (logarithmically)
with increasing $N$. This slow growth is a consequence of extreme value statistics
and has 
interesting implications in
ecological context in estimating the home range
of a herd of animals with population size $N$.
\end{abstract}

\pacs{05.40.-a, 02.50.-r, 87.23.Cc}

\maketitle

\begin{figure}
\begin{center}
\subfigure[]{\includegraphics[height=3cm, width=4cm]{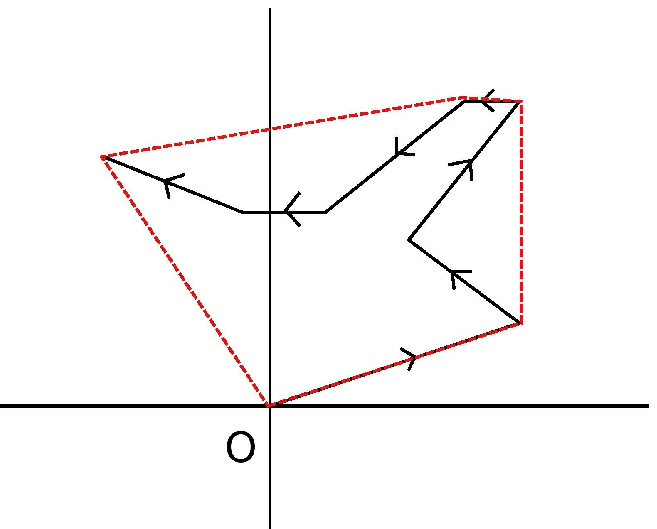}\label{GWCH}}
\subfigure[]{\includegraphics[height=3cm, 
width=4cm]{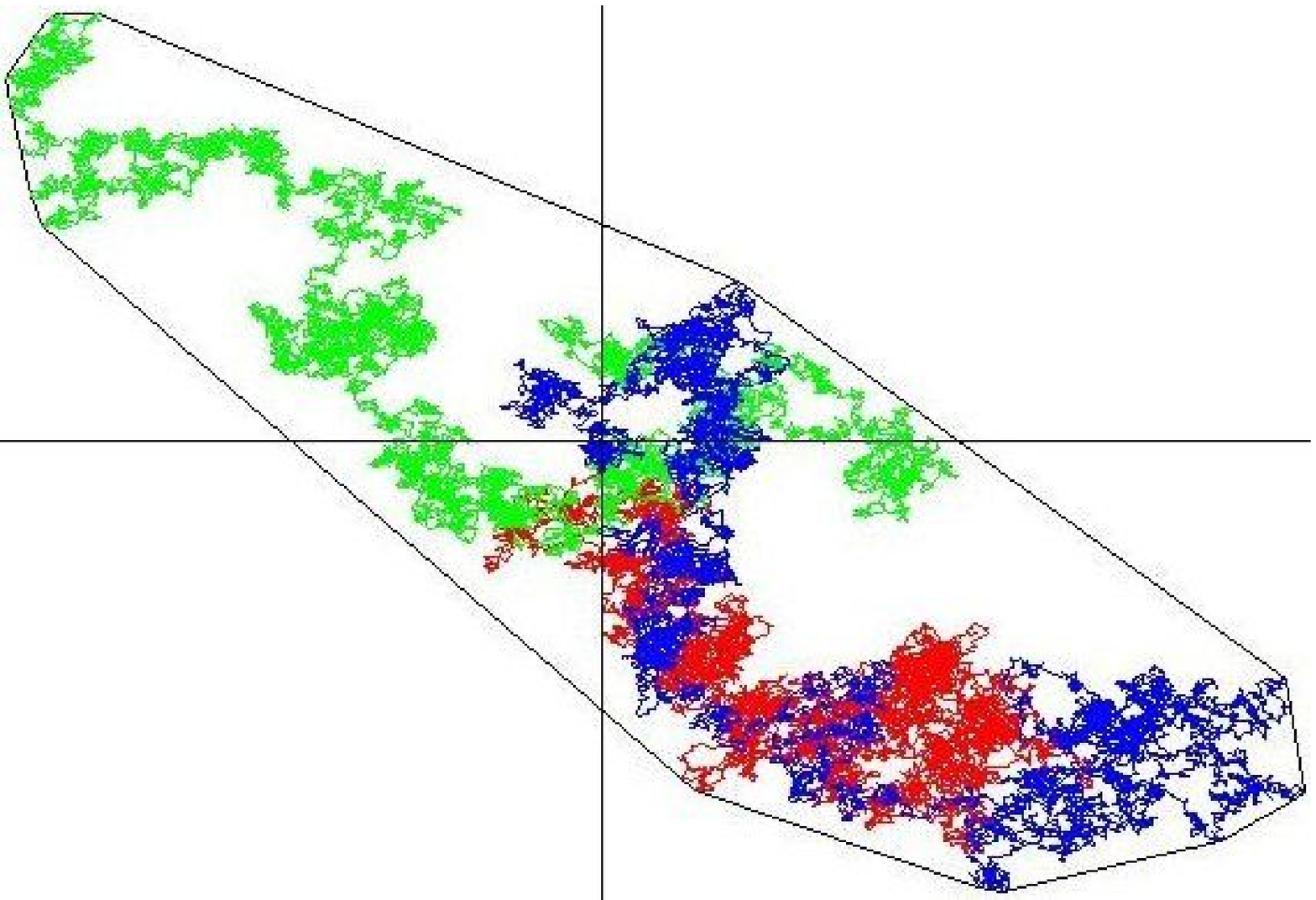}\label{CHBB}}
\caption{Convex hull of  (a) a 7-step random walk starting at $O$ (b)  
3~closed Brownian paths.}
\end{center}
\end{figure}

Ecologists often need to 
estimate the home range of an animal or a group of animals, 
in particular for habitat-conservation planning \cite{ConsPlan}. 
Home range of a group of animals simply means the two dimensional
space over which they typically 
move around in search of food. There 
exist various methods to estimate this home range, based on the monitoring 
of the positions of the animals over a certain period of time~\cite{Worton}. One method 
consists in drawing the minimum convex polygon 
enclosing all monitored positions, called the convex hull. While this may seem 
simple minded, it 
remains, under certain circumstances, the best way to proceed~\cite{Folia}. The monitored 
positions, for one animal, will appear as the 
vertices of a path whose statistical properties will depend on the type 
of motion the animal is performing. In particular, during phases of food 
searching known as foraging, the monitored positions can be described as 
the vertices of a random walk in the plane~\cite{ERW,BRW}. 
For 
animals whose daily motion consists mainly in foraging, quantities of 
interest about their home range, such as its perimeter and area, can be 
estimated through the average perimeter and area of the convex hull of the 
corresponding random walk (Fig.~\ref{GWCH}). If the recorded positions are 
numerous (which might result from a very fine and/or long monitoring), the 
number of steps of the random walker becomes large and
to a good approximation the trajectory of a discrete-time planar random walk 
(with finite variance of the step sizes) can
be replaced by a continuous-time planar Brownian motion of a certain duration $T$. 

The home range of a single animal can thus be characterized by
the mean perimeter and area of the convex hull of a planar Brownian motion 
of duration $T$ starting at origin $O$.
Both `open' (where the endpoint of the path
is free) and `closed' paths (that are constrained to
return to the origin in time $T$) are of interest. The latter corresponds, for instance,
to an animal returning
every night to its nest after spending the day foraging in the
surroundings. For an `open' path, the mean perimeter $\langle L_1 \rangle = 
\sqrt{8 \pi T}$
and the mean area $\langle A_1 \rangle = \pi T/2 $ are known
in the mathematics literature~\cite{Ta,LetacFS,El}. For a  `closed' path, only 
the mean
perimeter is known~\cite{Go}: $\langle L_1 \rangle =\sqrt{\pi ^3 T/2}$.

In any given habitat, an animal is however hardly alone but they live in herds
with typically a large population size $N$. To study their global home range via
the convex hull model, one needs to study the convex hull of $N$ planar
Brownian motions. Assuming that the animals do not develop any substantial
interaction between them during foraging, this leads us to the main question
addressed in this Letter: what is the mean perimeter and area of the convex hull
of $N$ independent planar Brownian motions (both open and closed) each of duration $T$?
This question is of vital importance in ecological conservation: if the population size 
increases by, say
ten-fold, by how much should one increase the home range, i.e., the conservation area?     

Using the standard scaling property of Brownian motion, ${\rm length scale} \sim 
({\rm time scale})^{1/2}$, it follows that the 
mean perimeter and area of the global convex hull of $N$ independent 
Brownian paths will behave as $\langle L_N \rangle = \alpha _N \sqrt{T}$ and 
$\langle A_N \rangle = \beta _N T$ for all $T$. The main challenge is to estimate
the $N$-dependence of the prefactors $\alpha_N$ and $\beta_N$. The central
result of this Letter is to provide exact formulae for $\alpha_N$ and $\beta_N$ for all
$N$, both for open and closed paths. 

%For example, for open paths, 
%we recover the results for $N=1$, but in addition we get $\alpha_2= 
%4\sqrt{\pi}=7.08982\ldots$ and $\beta_2=\pi=3.14158\ldots$ ($N=2$), $\alpha_3= 24 {\rm 
%tan}^{-1}(1/\sqrt{2})/\sqrt{\pi}=8.33393\ldots$ and $\beta_3= 
%\pi+3-\sqrt{3}=4.40954\ldots$ ($N=3$) etc.  

Most interestingly we find that
the mean perimeter and area of the convex hull increases very slowly with $N$ for large 
$N$:
$\alpha _N
\simeq 2\pi \sqrt{2\ln N}$, $\beta _N \simeq 2\pi \ln N$ for open paths,
and $\alpha _N (c)\simeq  \pi \sqrt{2\ln N}$, $\beta _N(c) \simeq \frac{\pi}{2}
\ln N$ for closed paths, $c$ referring to closed paths. 
This leads to our main conclusion: the home range increases very slowly with 
increasing population size.
Indeed, the $\ln N$ behaviour of the prefactor $\beta_N(c)$
in the mean area indicates that, for instance, a $10$-fold increase in the 
number of animals in the herd will result, on average, in the addition of 
only about $3.6\ T$ units of area to the home range of the herd ---~a good 
news for conservation.  

To proceed, let $I$ denote any set of points on the plane with coordinates 
$(x_i,y_i)$. Let $C$ denote the convex hull of $I$, i.e., the minimal
convex polygon enclosing this set. 
\begin{figure}[h]
\begin{center}
\includegraphics[height=5cm, width=6cm]{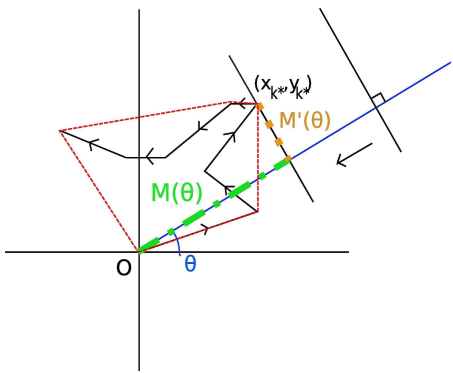}
\end{center}
\caption{Support function of a 7-step random walk}\label{SupFct}
\end{figure}
Geometrically, the convex hull can be constructed very simply:
consider any arbitrary direction from
the origin $O$ specified by the angle $\theta$ with respect to the $x$ axis.
Bring a straight line 
from infinity perpendicularly along direction $\theta$ and
stop when it touches a point of the set $I$ [e.g., in Fig.~\ref{SupFct}, this
point has co-ordinates $(x_{k^*},y_{k^*})$].
Let $M(\theta)$ denote the Euclidean distance of this perpendicular
line from the origin when it stops, measuring the maximal
extension of the set $I$ along the direction $\theta$. 
This support function can be simply written as
\begin{equation}
M(\theta)= \max _{i \in I} \left\{x_i \cos \theta +y_i
\sin\theta\right\}.
\label{M}
\end{equation}
Knowing $M(\theta)$, one can express the perimeter and the area 
of the convex hull $C$ via Cauchy's formulae~\cite{Cau}:
\begin{eqnarray}
L &=&\int_{0}^{2\pi}d\theta  M(\theta) \label{C1}\\
A &=&\frac{1}{2}\int_{0}^{2\pi}d\theta \left( M^2 (\theta) - 
\left(M'(\theta)\right)^2\right), \label{C2}
\end{eqnarray}
where $M'(\theta)= dM/d\theta$.
In this Letter we show how these two formulae can be
used to compute the
mean perimeter and area of the convex hull of $N$ planar
Brownian motions.

To illustrate the main idea let us start with a single open Brownian path, the 
generalizations
to $N>1$ and for closed paths will follow. In this case the set $I$ consists of the 
vertices
of a Brownian path of duration $T$ starting at the origin $O$. Since it is a continuous 
path, we
can conveniently label the coordinates by $\mathcal{B}(\tau)=(x(\tau),y(\tau))$
with
$0\le \tau\le T$. Here $x(\tau)$ and $y(\tau)$ are just two
independent one dimensional Brownian paths each of duration $T$ and
evolve via the Langevin equations: ${\dot x}(\tau)= \eta_x(\tau)$
and ${\dot y}(\tau)=\eta_y(\tau)$ where $\eta_x(\tau)$ and $\eta_y(\tau)$
are independent Gaussian white noises, each with zero mean and
delta correlated, e.g.,  $\langle \eta_x(\tau)\eta_x(\tau')\rangle 
=\delta(\tau-\tau')$ and the same for $\eta_y$. Clearly then $\langle x^2(\tau)\rangle=\tau$
and $\langle y^2(\tau)\rangle =\tau$.

\begin{figure}
\begin{center}
\subfigure[]{\includegraphics[height=3.6cm, width=4.25cm]{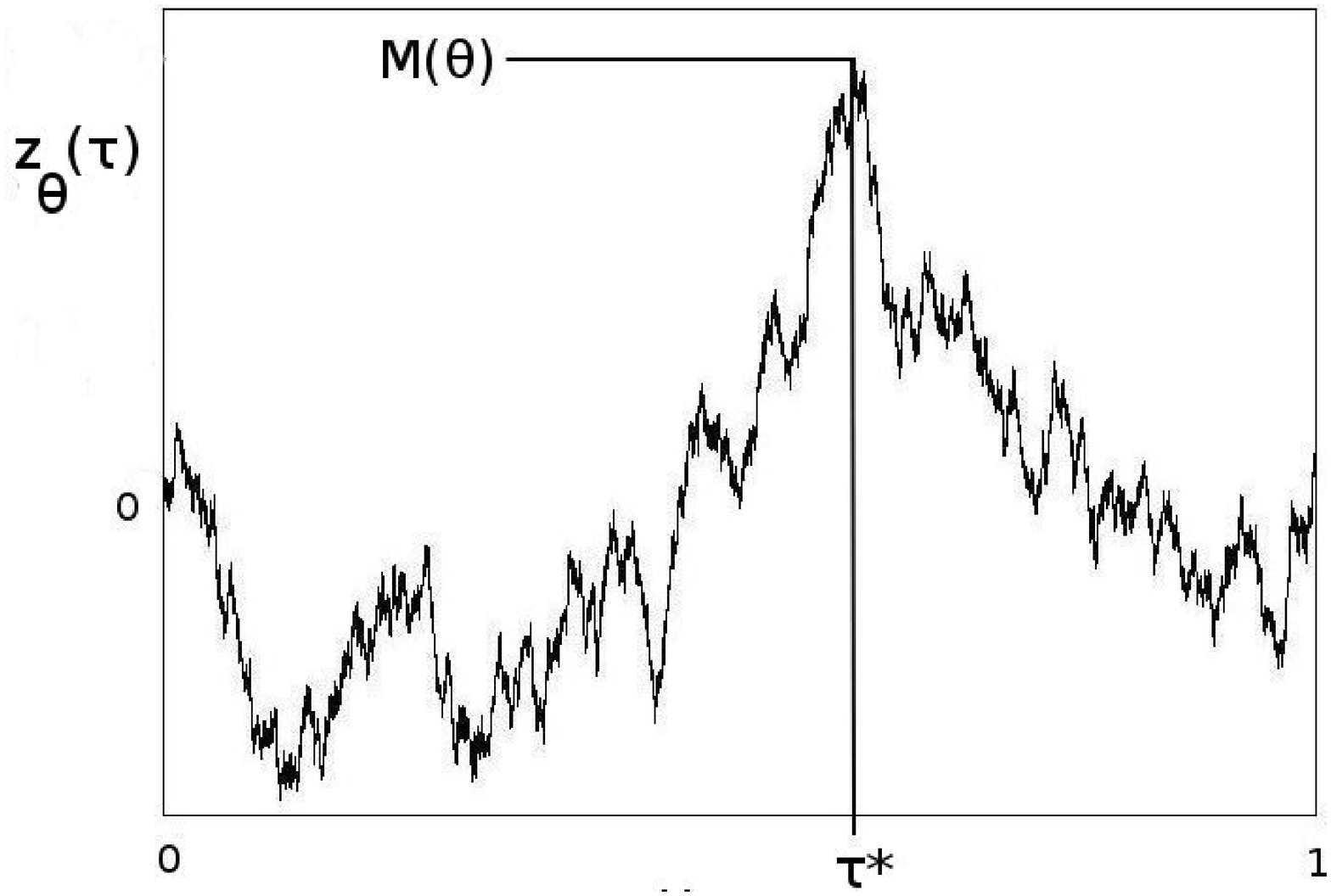}\label{z}}
\subfigure[]{\includegraphics[height=3.5cm, width=4.25cm]{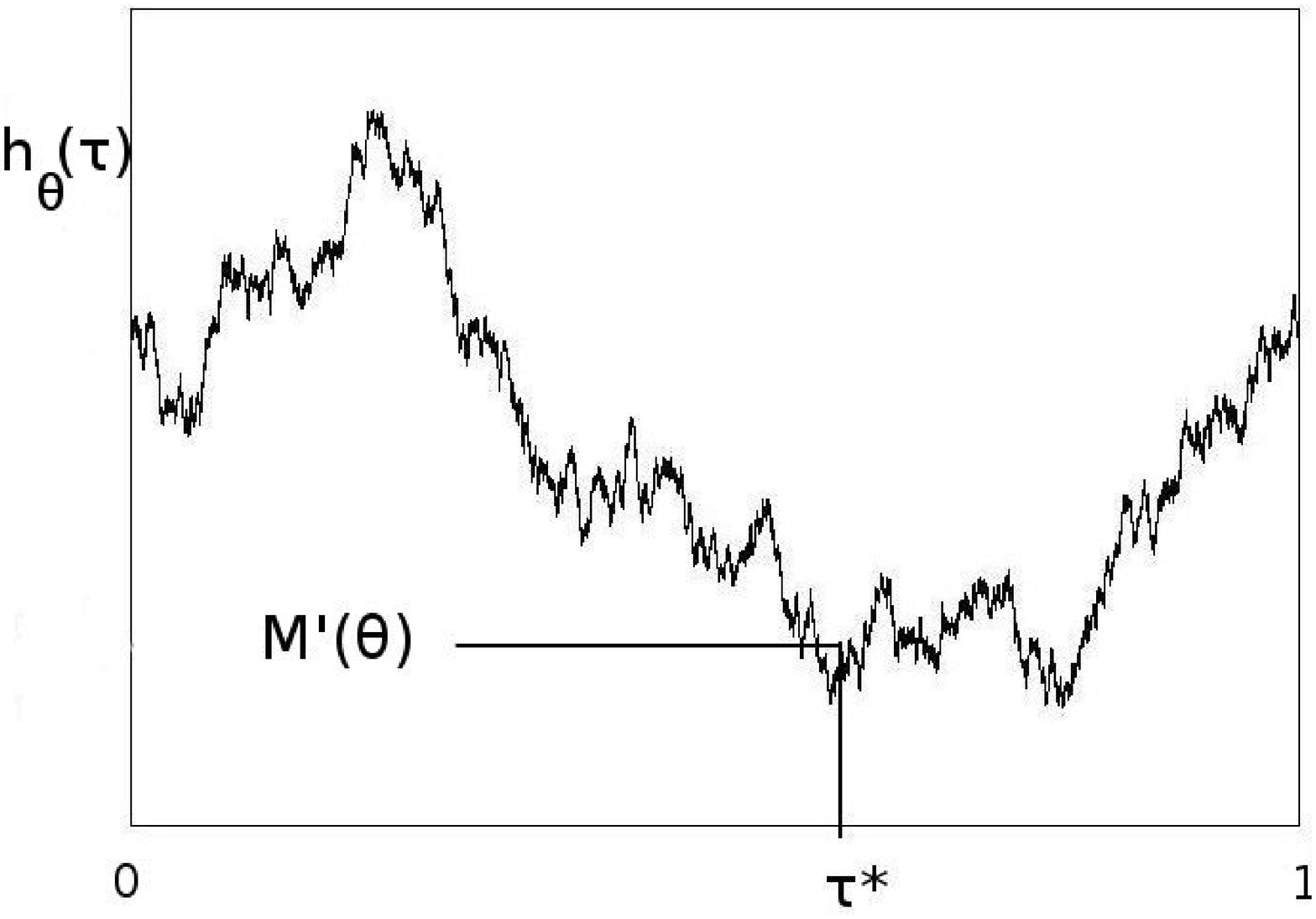}\label{h}}
\caption{(a) Location $\tau^*$ of the maximum $M(\theta)$ of $z_\theta(\tau)$  and (b) 
corresponding value of $M'(\theta)=h_\theta(\tau^*)$}
\end{center}
\end{figure}

Let us now consider a fixed direction $\theta$. Then, $z_\theta(\tau)= x(\tau)\cos \theta
+y(\tau) \sin \theta$ and $h_\theta(\tau)= -x(\tau) \sin \theta
+y(\tau)\cos\theta$ are also two independent one dimensional 
Brownian motions (each of duration $T$)
parametrized by $\theta$.  
Then $M(\theta)$ in Eq. (\ref{M}) is simply
the maximum of the one dimensional Brownian motion $z_{\theta}(\tau)$ over
the time interval $\tau\in [0,T]$, i.e., $M(\theta)= {\rm max}[ z_{\theta}(\tau)]$.  
Let $\tau^*$ be the time at which this maximum is achieved.
Then, $M(\theta)= z_\theta(\tau^*)= x(\tau^*) \cos \theta+ y(\tau^*) \sin \theta$.
Taking derivative with respect to $\theta$ gives  
$M'(\theta)= -x(\tau^*) \sin \theta+ y(\tau^*) \cos \theta= h_{\theta}(\tau^*)$.
Thus, while $M(\theta)$ is the maximum value of the first 
Brownian motion 
$z_{\theta}(\tau)$, $M'(\theta)$ is the value of the second independent Brownian motion
$h_\theta(\tau)$ but at the time $\tau=\tau^*$ at which the first one achieves its maximum 
(see 
Fig.~\ref{z} and
\ref{h}). 
Note, in particular, that for $\theta=0$, $z_0(\tau)=x(\tau)$
and $h_0(\tau)= y(\tau)$. Thus $M(0)$ is the maximum of $x(\tau)$ in $\tau\in [0,T]$
while $M'(0)=y(\tau^*)$ is the value of $y$ at the time $\tau^*$ when $x$ achieves
its maximum. 

Taking expectation and using isotropy
of the mean over the choice of directions, it follows that
\begin{eqnarray}
\langle L\rangle &=& 2\pi \ \langle M(0) \rangle \label{IAC1}\\
\langle A\rangle &=& \pi \left( \langle [M(0)]^2 \rangle - \langle
[M'(0)]^2\rangle\right). \label{IAC2}
\end{eqnarray}
For a one dimensional Brownian 
motion $x(\tau)$ over $[0,T]$, the distribution of its maximum is well known~\cite{Feller}.
The cumulative distribution, $Q_1(m,T)={\rm Prob}[M(0)\le m]={\rm erf}(m/\sqrt{2T})$, where
${\rm erf}(z)= \frac{2}{\sqrt{\pi}}\,\int_0^z e^{-u^2}\, du$.
The first two moments can be easily computed: 
$\langle M(0)\rangle = 
\sqrt{2T/\pi}$ and $\langle [M(0)]^2\rangle = T$.
Eq. (\ref{IAC1}) then gives the mean perimeter, $\langle 
L_1\rangle=\alpha_1 \sqrt{T}$
with $\alpha_1= \sqrt{8\pi}$.  
The calculation of the mean area is slightly trickier since we need $\langle 
[M'(0)]^2\rangle$.  
We first note that for any fixed time $\tau^*$, $E [y^2(\tau^*)] = \tau^*$
(since it is a free Brownian motion) where the expectation is taken over
all realizations of the process $y$ at fixed $\tau^*$. But $\tau^*$ itself
is a random variable, being the time at which the first process $x$ achieves its maximum.
Its distribution is also well known and given by the celebrated arcsine law of 
L\'evy~\cite{Levy}:
$P(\tau^*, T) = [\tau^* (T-\tau^*)]^{-1/2}/\pi$. Thus, averaging further over
$\tau^*$ taken from this distribution, we finally get
$\langle [M'(0)]^2\rangle= \langle \tau^*\rangle = T/2$. 
Eq. (\ref{IAC2}) then gives the exact mean area $\langle A_1\rangle = \beta_1 T$
with $\beta_1= \pi/2$.

For a single closed path, the analysis is similar, except that $x(\tau)$
and $y(\tau)$ are now two independent one dimensional Brownian bridges
of duration $T$, i.e., both start at the origin but are constrained
to return to the origin at time $T$: $x(0)=x(T)=0$ and $y(0)=y(T)=0$. The distribution of the 
maximum
of a Brownian bridge is also well known~\cite{Feller} and its first two moments are
given by: $\langle M(0)\rangle =
\sqrt{\pi T/8}$ and $\langle [M(0)]^2\rangle = T/2$. Thus, the mean perimeter
of the convex hull is given from Eq. (\ref{IAC1}): $\langle
L_1\rangle=\alpha_1(c) \sqrt{T}$
with $\alpha_1(c)= \sqrt{\pi^3/2}$. To compute the mean area, we note
that for a Brownian bridge $y(\tau)$, at fixed time $\tau^*$, $E [y^2(\tau^*)] = 
\tau^*(T-\tau^*)/T$. Moreover, the distribution of $\tau^*$, the time at which
a bridge achieves its maximum is known to be uniform~\cite{Feller}. Thus, taking average 
over $\tau^*$
with uniform distribution, $P(\tau^*,T)=1/T$, we get  
$\langle M'(0)^2\rangle= \langle y^2(\tau^*)\rangle = T/6$.
Eq. (\ref{IAC2}) then gives the exact mean area of the convex hull
of a planar closed Brownian path: $\langle A_1\rangle = \beta_1(c)\, T$
with $\beta_1(c)= \pi/3$. To our knowledge, this result, even for a single closed path,
appears to be new.

This method can then be generalized to $N$ independent
planar Brownian paths, open or closed. We now have two sets of
$N$ Brownian paths: $x_j(\tau)$ and $y_j(\tau)$ ($j=1,2,\ldots,N$).
All paths are independent of each other. 
Since isotropy holds,
we can still use Eqs. (\ref{IAC1}) and (\ref{IAC2}), except
that $M(0)$ now denotes the global maximum of a set of $N$
independent one dimensional Brownian paths (or bridges for closed paths)
$x_j(\tau)$ ($j=1,2,\ldots,N$), each of duration $T$, 
\begin{equation}
M(0)= \max_{1\le j\le N}\,\max_{0\le \tau\le T}\left[x_1(\tau),x_2(\tau),\ldots, 
x_N(\tau)\right].
\label{MN1}
\end{equation}
Let $j_*$ and $\tau^*$ denote the label of the path and the time at which this
global maximum is achieved. Then, using argument similar to the $N=1$ case, it is easy to 
see that $M'(0)= y_{j_*}(\tau^*)$, i.e., the position of the $j_*$-th $y$ path
at the time when the $x$ paths achieve their global maximum.

To compute the first two moments of $M(0)$, we first compute the distribution
$P_N[M(0),T]$ of the global maximum of $N$ independent Brownian paths (or 
bridges) $x_j(\tau)$. This is a standard extreme value calculation. 
Consider first $N$ open Brownian paths.
It is easier to compute the cumulative probability, $Q_N(m,T)= {\rm Prob}[M(0)\le 
m]$. Since the Brownian paths are independent, it follows that $Q_N(m,T)= [Q_1(m,T)]^N$,
where $Q_1(m,T)= {\rm erf}(m/\sqrt{2T})$ for a single path mentioned before.
Knowing this cumulative distribution
$Q_N(M(0),T)$, the first two moments $\langle M(0)\rangle$ and $\langle [M(0)]^2\rangle$
can be computed for all $N$. Using the result for $\langle M(0)\rangle$ in Eq. 
(\ref{IAC1}) gives us the mean perimeter, $\langle L_N\rangle =\alpha_N \sqrt{T}$
with
\begin{equation}
\alpha_N = 4 N \sqrt{2\pi} \int _0^\infty du\ u\ e^{-u^2}
\left[\text{erf}\left(u\right)\right]^{N-1}.
\label{R1}  
\end{equation}
The first few values are: $\alpha_1= \sqrt{8\pi}=5.013..$, $\alpha_2= 
4\sqrt{\pi}=7.089..$,
$\alpha_3= 24\,{\rm tan}^{-1}(1/\sqrt{2})/\sqrt{\pi}=8.333..$ etc. (see Fig. 4 for
a plot of $\alpha_N$ vs. $N$).
For large $N$, one can
analyse the integral in Eq.~(\ref{R1}) by the saddle point method giving~\cite{details},
$\alpha_N \simeq 2\pi \sqrt{2 \ln N}$. 
This logarithmic dependence on $N$
is thus a direct consequence of extreme value statistics~\cite{Gumbel}
and the calculation of the mean perimeter
of the convex hull of $N$ paths is a nice application of the extreme value 
statistics.

To compute the mean area, we need to calculate $\langle [M'(0)]^2\rangle$ in
Eq.~(\ref{IAC2}). We proceed as in the $N=1$ case. For a fixed label $j$ and fixed 
time $\tau$, the expectation $E[y_j^2(\tau)]=\tau$ which follows
from the fact that $y_j(\tau)$ is simply a Brownian motion.
Thus, $E[y_{j_*}^2(\tau^*)]=\tau^*$. Next, we need to average over
$\tau^*$ which is the time at which the global maximum in Eq. (\ref{MN1})
happens. The distribution $P_N(\tau^*,T)$ of the time of global
maximum of $N$ independent Brownian motions, to our knowledge,
is not known in the probability literature. We were able to compute
this exactly for all $N$~\cite{details}. Skipping details, we find
that $P_N(\tau^*,T)= \frac{1}{T}\,F_N(\tau^*/T)$ where 
\begin{equation}
F_N(z)= \frac{a_N}{\sqrt{z(1-z)}}\,\int_0^{\infty}dx\, x\, e^{-x^2}\, 
\left[{\rm erf}\left(x\sqrt{z}\right)\right]^{N-1}
\label{maxtN}
\end{equation}  
and $a_N$ is a normalization constant fixed by, $\int_0^1 
F_N(z)\,dz=1$.
It is easy to check that for $N=1$, it reproduces the arcsine law mentioned before.
Averaging over $\tau^*$ drawn from this distribution, we can then
compute $\langle [M'(0)]^2\rangle = \int_0^{T} \tau^* P_N(\tau^*,T)\, d\tau^*$.
Substituting this in Eq. (\ref{IAC2}) gives the exact mean area 
for all $N$, $\langle A_N \rangle =\beta_N T$ with
\begin{equation}
\beta_N= {4N}\,{\sqrt{\pi }}\, \int _0 ^\infty du\ u\
\left[\text{erf}(u)\right]^{N-1}\left(u e^{-u^2}-g(u)\right)
\label{R3}
\end{equation}
where $g(u)=\frac{1}{2\sqrt{\pi}}\int_0^1 \frac{e^{-{u^2}/{t}}\
dt}{\sqrt{t(1-t)}}$. For example, the first few values are given by,
$\beta_1= \pi/2=1.570..$, $\beta_2= \pi= 3.141..$, $\beta_3= \pi+3-\sqrt{3}= 4.409..$
etc. (Fig. 4 shows a plot of $\beta_N$ vs $N$). The large $N$ analysis 
gives~\cite{details}, $\beta_N \simeq 2\pi \ln N$. Thus for large $N$,
the shape of the convex hull approaches a circle of radius $\sqrt{2\ln N}$ which,
incidentally, coincides with the set of distinct sites
visited by $N$ Brownian motions~\cite{larralde}.

For $N$ closed Brownian planar paths one proceeds in a similar way.
Without repeating the analysis, we just mention our main results~\cite{details}. The
mean perimeter and area are given by, $\langle L_N\rangle =\alpha_N(c) 
\sqrt{T}$ and $\langle A_N\rangle =\beta_N(c) T$ where, for all $N$, 
\begin{eqnarray}
\alpha_N(c)&= & \frac{\pi ^{3/2}}{\sqrt{2}}\sum
_{k=1}^{N}\binom{N}{k}\frac{(-1)^{k+1}}{\sqrt{k}} \label{R2} \\
\beta_N(c) &=& \frac{\pi}{2}\left[\sum
_{k=1}^{N}\frac{1}{k}-\frac{N}{3}+\frac{1}{2}\sum_{k=2}^{N}(-1)^k\, f(k)\right]\label{R4}
\end{eqnarray}
and $f(k)=\binom{N}{k}\,(k-1)^{-3/2}\left(
k\tan^{-1}(\sqrt{k-1})-\sqrt{k-1}\right)$.
The first few values are: $\alpha_1(c)= \sqrt{\pi^3/2}=3.937..$,
$\alpha_2(c)= \sqrt{\pi^3}(\sqrt{2}-1/2)=5.090..$, $\alpha_3(c)= 
\sqrt{\pi^3}(3/\sqrt{2}-3/2+1/\sqrt{6})=5.732..$ and 
$\beta_1(c)=\pi/3=1.047..$, $\beta_2(c)= \pi(4+3\pi)/24=1.757..$,
$\beta_3(c)= 2.250..$ etc. (see Fig. 4 for a plot of $\alpha_N(c)$
and $\beta_N(c)$ vs. $N$).
Large $N$ analysis shows that~\cite{details}, $\alpha_N(c)\simeq 
\pi \sqrt{2\ln N}$ and $\beta_N(c) \simeq \frac{\pi}{2}
\ln N$, smaller respectively by a factor $1/2$ and $1/4$ than
the corresponding results for open paths.

\begin{figure}
\begin{center}
\includegraphics[height=7cm, width=8cm]{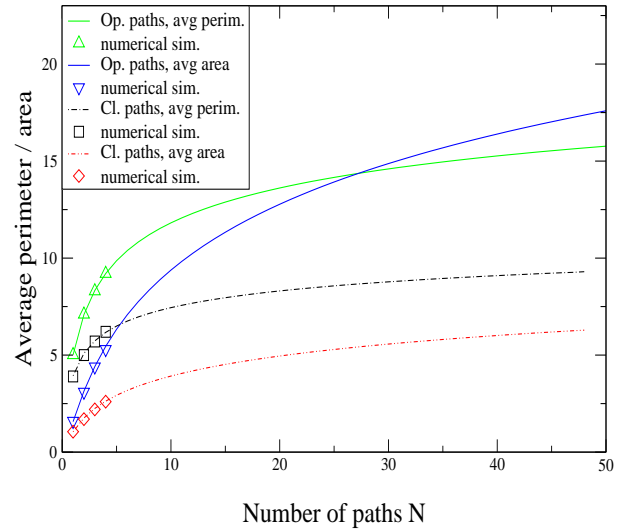}
\end{center}
\caption{Setting $T=1$, the analytical results for average perimeter $\alpha_N$ [Eq. 
(\ref{R1})] and area $\beta_N$ [Eq. (\ref{R3})] of $N$ open (Op.) Brownian paths,
and similarly the average perimeter 
$\alpha_N(c)$ [Eq. (\ref{R2})] and area $\beta_N(c)$ [Eq. (\ref{R4})] of $N$
closed (Cl.) Brownian paths, plotted against $N$. The symbols denote results
from numerical simulations (up to $N=4$).}
\label{OP}
\end{figure}

We have also computed the mean perimeter and area of the convex hull
of $N=1,2,3,4$ Brownian paths (both open and closed) via
numerical simulations. 
For each $N$, we constructed $N$ independent normal Gaussian random
walks of $10^4$ steps each with a time step $\Delta \tau = 10^{-4}$. For each
realisation of the $N$ walks, we constructed the convex hull 
using the Graham scan algorithm~\cite{Graham} and computed its perimeter and area 
and then averaged over $10^3$ samples. We find excellent agreement with
our analytical predictions (Fig.~\ref{OP}).

The method presented here is general and can, in principle, be applied to
compute the mean perimeter and area of the convex hull of any
set of random points.
In fact, when the set consists of just random points, each drawn
independently from a given distribution, the statistics of the perimeter
and area of their convex hull has been studied extensively in
various different contexts~\cite{Renyi}. Our method, using extreme
value statistics, can easily recover these results~\cite{details}, but
can go further e.g., when the points are correlated as in the case
of Brownian paths. 

This work leads to several interesting open questions. 
It would be interesting to extend the results presented here for normal
diffusion to the case
where the trajectories of animals such as birds, deers or spider monkeys undergo 
anomalous 
diffusion, e.g., 
L\'evy flights etc.~\cite{GM,KS,Ramos}. 
Another interesting question concerns the effect of interactions or collective
behavior of the animals on the statistics of their convex hulls. For animals
like birds that move in $3$-dimensional space, it would be interesting
to study the statistics of the convex polytope of their trajectories, such
as the mean surface area and the volume of such a polytope~\cite{details}.
Finally, the distribution of the time of maximum $\tau^*$, a crucial ingredient
in our method, has recently
been computed exactly for a variety of constrained one dimensional Brownian 
motions~\cite{julien}. These results
may be useful to study
the statistics of convex hulls of constrained planar Brownian paths (see e.g. 
~\cite{BL}).

We thank D. Dhar and H. Larralde for useful 
discussions.

\end{document}